\documentclass[numreferences]{kluwer}

\usepackage[dvips]{graphicx}

\begin{document}

\typeout{^^J*** Kluwer Academic Publishers - prepress department^^J***
    Documentation for book editors using the Kluwer class files}
\begin{article}
\begin{opening}
\title{Quantum computing with many superconducting qubits}
\runningauthor{Kluwer Academic Publishers}
\runningtitle{Editor's Manual}
\author{J. Q. You$^1$, J. S. Tsai$^{1,2}$ and
Franco Nori$^{1,3,}$\thanks{Corresponding author
(e-mail address:~nori@umich.edu)}}
\institute{$^1$Frontier Research System, The Institute of Physical
and Chemical Research (RIKEN), Wako-shi 351-0198, Japan\\
$^2$NEC Fundamental Research Laboratories, Tsukuba,
Ibaraki 305-8051, Japan$^{\dag}$\\
$^3$Center for Theoretical Physics, Physics Department,
Center for the Study of Complex Systems, The University of
Michigan, Ann Arbor, MI 48109-1120, USA\thanks{Permanent address.}}

\begin{abstract}
Two of the major obstacles to achieve quantum computing (QC) are
(i)~scalability to many qubits and (ii)~controlled connectivity
between any selected qubits.  Using Josephson charge qubits, here
we propose an experimentally realizable method to efficiently
solve these two central problems.
Since any two charge qubits can be effectively coupled by an
experimentally accessible inductance, the proposed QC architecture
is {\it scalable}.
In addition, we formulate an efficient and realizable QC scheme
that requires {\it only one} (instead of two or more) two-bit
operation to implement conditional gates.
\end{abstract}
\end{opening}

\section*{Introduction}

Josephson-qubit devices~\cite{MSS} are based on the charge and phase degrees
of freedom. The charge qubit is achieved in a Cooper-pair box~\cite{NPT},
where two dominant charge states are coupled through coherent Cooper-pair
tunneling~\cite{ASZ}.
% while the phase qubit is based on two different
%flux states in a small superconducting-quantum-interference-device
%(SQUID) loop~\cite{MOOIJ,ORLANDO}.
%
%
Using Cooper-pair tunneling in Josephson charge
devices~\cite{NCT,BOUCH} and via spectroscopic measurements for
the Josephson phase device~\cite{VAL,FRIED}, it has been possible
to experimentally observe energy-level splitting and related
properties for state superpositions.
In addition, using Josephson charge devices prepared in a
superposition of two charge states~\cite{NPT}, coherent
oscillations were observed.
While operating at the degeneracy point, the charge-qubit states
are highly coherent~\cite{VION} ($Q=2.5\times 10^4$), with a
decoherence time of $\tau\sim 500$~ns.
These important experimental results indicate that the Josephson
charge and phase devices are potentially useful for solid-state
qubits in quantum information processing.
Important open problems would now include implementing a {\it
two-bit coupling\/} and then {\it scaling up\/} the architecture
to many qubits.
Here, we propose a new quantum-computing (QC) scheme based on
scalable charge-qubit structures.
We focus on the Josephson charge qubit realized in a Cooper-pair
box.

\subsection*{Coupling qubits}

The Coulomb interaction between charges on different islands of
the charge qubits would seem to provide a natural way of coupling
Josephson charge qubits (e.g., to connect two Cooper-pair boxes
via a capacitor).
Using this type of capacitive interbit coupling, a two-bit
operation~\cite{PFP} similar to the controlled-NOT gate was
derived.
However, as pointed out in \cite{MSS}, it is difficult in this
scheme to switch on and off the coupling.  Also, it is hard to
make the system scalable because only neighboring qubits can
interact.
Moreover, implementations of quantum algorithms such as the
Deutsch and Bernstein-Vazirani algorithms were studied using a
system of Josephson charge qubits \cite{SIEWERT}, where it was
proposed that the nearest-neighbor charge qubits would be coupled
by tunable dc SQUIDs.
In the semiconductor literature, scalability often refers to 
reducing the size of the device (packing more components). 
In QC, scalability refers to increasing the number of qubits
coupled with each other.

A suggestion for a scalable coupling of Josephson charge qubits
was proposed~\cite{MSS,ASZ} using oscillator modes in a $LC$
circuit formed by an inductance and the qubit capacitors.
In this proposal, the interbit coupling can be switched and any
two charge qubits could be coupled.
Nevertheless, there is no efficient (that is, using one two-bit
operation) QC scheme for this proposal~\cite{MSS,ASZ} in order to
achieve conditional gates---e.g., the controlled-phase-shift and
controlled-NOT gates.
In addition, the calculated interbit coupling terms~\cite{MSS,ASZ}
only apply to the case when the following two conditions are met:

(i)~The quantum manipulation frequencies, which are fixed
experimentally, are required to be much smaller than the
eigenfrequency $\omega_{LC}$ of the $LC$ circuit. This condition
{\it limits} the allowed number $N$ of the qubits in the circuit
because $\omega_{LC}$ scales with $1/\sqrt{N}$.  In other words,
the circuits in \cite{MSS,ASZ} are not really scalable.

(ii)~The phase conjugate to the total
charge on the qubit capacitors fluctuates weakly.

\subsection*{Improved and scalable coupling between any selected
qubits}

The limitations listed above do not apply to our approach.
In our scheme, a common inductance, but no $LC$ circuit, is used
to couple all Josephson charge qubits.
In our proposal, both dc and ac supercurrents can flow through the
inductance, while in~\cite{MSS,ASZ} only ac supercurrents can flow
through the inductance and it is the $LC$-oscillator mode that
couples the charge qubits.
These yield different interbit couplings (e.g., $\sigma_y
\sigma_y$ type~\cite{MSS,ASZ} as opposed to $\sigma_x \sigma_x$ in
our proposal).

We employ two dc SQUIDs to connect each Cooper-pair box in order
to achieve a {\it controllable interbit coupling}.
Our proposed QC architecture is scalable in the sense that {\it
any \/} two charge qubits ({\it not \/} necessarily neighbors) can
be effectively coupled by an experimentally accessible inductance.
We also formulate \cite{YTN} an efficient QC scheme that requires
only one (instead of two or more) two-bit operation to implement conditional gates. \\

This Erice summer-school presentation is based on our work in \cite{YTN}.
Additional work on decoherence and noise-related issues appears in, e.g.,
\cite{PALA,YHN}. Also, work more focused on entanglement and readout issues
appears in \cite{YOU}. Other interesting studies on charge qubits can be
found in \cite{AV} for the adiabatic controlled-NOT gate, in \cite{FALCI}
for geometric phases, and in \cite{HEK,BRU,ARM,YN} for the
dynamics of a Josephson charge qubit coupled to a quantum resonator.

\section*{Proposed scalable and switchable quantum computer}

\begin{figure}
\centerline{\includegraphics[width=4.3in,
bbllx=45,bblly=410,bburx=540,bbury=670]{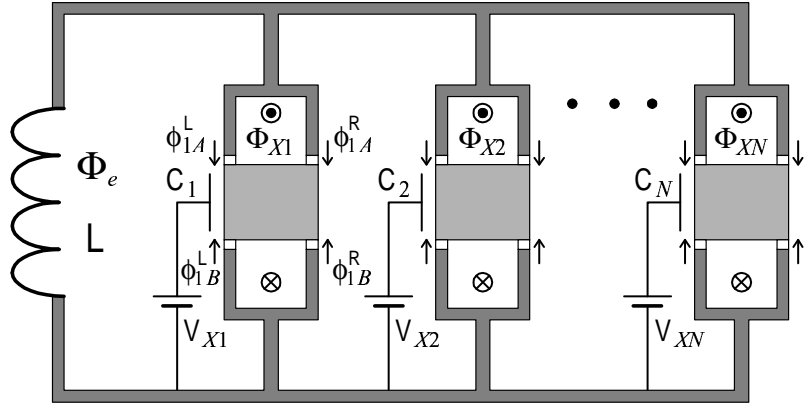}}
\caption{Schematic diagram of the proposed scalable and switchable
quantum computer.
Here, each Cooper-pair box is operated in the charging regime,
$E_{ck}\gg E^0_{Jk}$, and at low temperatures $k_B T \ll E_{ck}$.
Also, the superconducting gap is larger than $E_{ck}$, so that
quasiparticle tunneling is strongly suppressed.
All Josephson charge-qubit structures are coupled by a common
superconducting inductance.
 } \label{fig1}
\end{figure}

Figure 1 shows a proposed QC circuit consisting of $N$ Cooper-pair
boxes coupled by a common superconducting inductance $L$.
For the $k$th Cooper-pair box, a superconducting island with
charge $Q_k=2en_k$ is weakly coupled by two symmetric dc SQUIDs
and biased, through a gate capacitance $C_k$, by an applied
voltage $V_{Xk}$.
The two symmetric dc SQUIDs are assumed to be equal and all
Josephson junctions in them have Josephson coupling energy
$E^0_{Jk}$ and capacitance $C_{Jk}$.
The effective coupling energy is given by the SQUIDs, each one
enclosing a magnetic flux $\Phi_{Xk}$.  Each SQUID provides a
tunable coupling
$-E_{Jk}(\Phi_{Xk})\cos\phi_{kA(B)}$, with
\begin{equation}
E_{Jk}(\Phi_{Xk})=2E^0_{Jk}\cos(\pi\Phi_{Xk}/\Phi_0),
\end{equation}
and $\Phi_0=h/2e$ is the flux quantum.
The effective phase drop $\phi_{kA(B)}$,
with subscript $A(B)$ labelling the SQUID above (below) the island,
equals the average value,
$[\phi^L_{kA(B)}+\phi^R_{kA(B)}]/2$, of the phase drops
across the two Josephson junctions in the dc SQUID,
where the superscript $L$ ($R$) denotes the left (right) Josephson junction.
Above we have neglected the self-inductance effects of each SQUID
loop because the size of the loop is usually very small ($\sim 1$
$\mu$m).
The Hamiltonian of the system then becomes
\begin{equation}
H=\sum_{k=1}^N H_k+{1\over 2}LI^2 \, ,
\end{equation}
with $H_k$ given by
\begin{equation}
H_k=E_{ck}(n_k-n_{Xk})^2-E_{Jk}(\Phi_{Xk})(\cos\phi_{kA}+\cos\phi_{kB}).
\end{equation}
Here
\begin{equation}
E_{ck}=2e^2/(C_k+4C_{Jk})
\end{equation}
is the charging energy of the superconducting island and $I=\sum_{k=1}^NI_k$
is the total persistent current through the superconducting inductance,
as contributed by all coupled Cooper-pair boxes. %~\cite{standard}.
The offset charge $2en_{Xk}=C_kV_{Xk}$ is induced by the gate
voltage $V_{Xk}$.
The phase drops $\phi^L_{kA}$ and $\phi^L_{kB}$
are related to the total flux
\begin{equation}
\Phi=\Phi_e+LI
\end{equation}
through the inductance $L$ by the constraint
\begin{equation}
\phi^L_{kB}-\phi^L_{kA}=2\pi\Phi/\Phi_0,
\end{equation}
where $\Phi_e$ is
the externally applied magnetic flux threading the inductance $L$.
In order to obtain a simpler expression for the interbit coupling,
and without loss of generality, the magnetic fluxes through the
two SQUID loops of each Cooper-pair box are designed to have the
{\it same} values but {\it opposite} directions.
If this were not to be the case, the interbit coupling can still
be realized, but the Hamiltonian of the qubit circuits would just
take a more complicated form.
Because this pair of fluxes cancel each other in any loop
enclosing them, then
\begin{equation}
\phi^L_{kB}-\phi^L_{kA}=\phi^R_{kB}-\phi^R_{kA}.
\end{equation}
This imposes the constraint
\begin{equation}
\phi_{kB}-\phi_{kA}=2\pi\Phi/\Phi_0
\end{equation}
for the average phase drops across the Josephson junctions
in the SQUIDs.
The common superconducting inductance $L$ provides the coupling
among Cooper-pair boxes.
The coupling of selected Cooper-pair boxes can be implemented by
switching ``on" the SQUIDs connected to the chosen Cooper-pair
boxes.
In this case, the persistent currents through the inductance $L$
have contributions from all the coupled Cooper-pair boxes.
The essential features of our proposal can be best understood via
the very simplified diagram shown in Fig.~2.

\begin{figure}
\centerline{\includegraphics[width=4.7in,
bbllx=89,bblly=473,bburx=522,bbury=666]{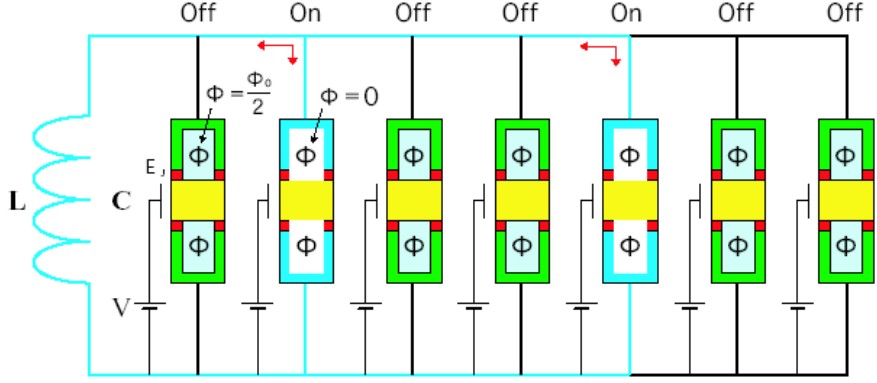}}
\caption{Simplified diagram of the circuit shown in Fig.~1. Here
we explicitly show how two charge qubits (not necessarily
neighbors) can be coupled by the inductance $L$, where the cyan
SQUIDs are switched on by setting the fluxes through the cyan
SQUID loops zero, and the green SQUIDs are turned off by choosing
the fluxes through the green SQUID loops as $\Phi_0/2$. This
applies to the case when any selected charge qubits are coupled by
the common inductance~\cite{web_color}.} \label{fig2}
\end{figure}

\subsection*{One-bit circuit}

As seen in Fig.~3(a), for any given Cooper-pair box, say $i$, when
\[
\Phi_{Xk}={1\over 2}\Phi_0, \ \ \ \ V_{Xk}=(2n_k+1)e/C_k
\]
for all boxes except $k=i$, the inductance $L$ only connects the
$i$th Cooper-pair box to form a superconducting loop.
The Hamiltonian of the system can be reduced to~\cite{YTN}
\begin{equation}
H=\varepsilon_i(V_{Xi})\,\sigma_z^{(i)}-\overline{E}_{Ji}(\Phi_{Xi},\Phi_e,L)
\;\sigma_x^{(i)},
\end{equation}
where
\begin{equation}
\varepsilon_i(V_{Xi})={1\over 2}E_{ci}[C_iV_{Xi}/e-(2n_i+1)]
\end{equation}
is controllable via the gate voltage $V_{Xi}$, while the intrabit coupling
$\overline{E}_{Ji}$
%(\Phi_{Xi},\Phi_e,L)$
can be controlled by both the applied external flux $\Phi_e$
through the common inductance, and the local flux $\Phi_{Xi}$
through the two SQUID loops of the $i$th Cooper-pair box.
Retained up to second-order terms in the expansion parameter
\begin{equation}
\eta_i=\pi LI_{ci}/\Phi_0,
\end{equation}
where
\begin{equation}
I_{ci}=-\pi E_{Ji}(\Phi_{Xi})/\Phi_0,
\end{equation}
we obtain
\begin{equation}
\overline{E}_{Ji}(\Phi_{Xi},\Phi_e,L)\,=
\,E_{Ji}(\Phi_{Xi})\cos(\pi\Phi_e/\Phi_0)\,\xi,
\end{equation}
with
\begin{equation}
\xi=1-{1\over 2}\eta_i^2\sin^2(\pi\Phi_e/\Phi_0).
\end{equation}
The intrabit coupling $\overline{E}_{Ji}$ in (9) is different from
that in~\cite{MSS,ASZ} because a very different contribution by
$L$ is considered.

\begin{figure}
\includegraphics[width=2in,height=1.3in,
bbllx=12,bblly=480,bburx=230,bbury=630]{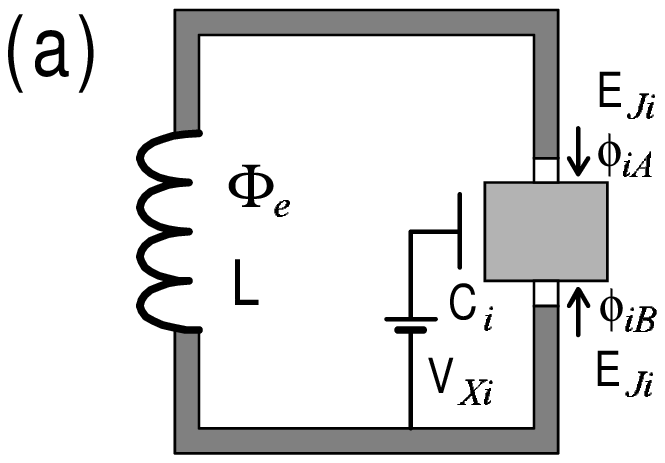}
\includegraphics[width=2.5in,height=1.33in,
bbllx=240,bblly=480,bburx=530,bbury=635]{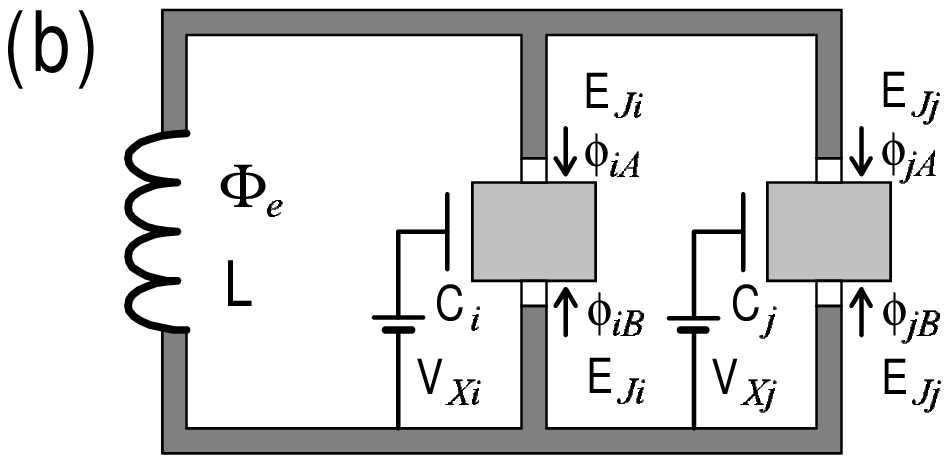}
\caption{(a) One-bit circuit with a Cooper-pair box connected to
the inductance. (b) Two-bit structure where two Cooper-pair boxes
are commonly connected to the inductance. Here, each SQUID
connecting the superconducting island is represented by an
effective Josephson junction.}
\label{fig3}
\end{figure}

\subsection*{Two-bit circuit}

To couple {\it any} two Cooper-pair boxes, say $i$ and $j$,
we choose
\[
\Phi_{Xk}={1\over 2}\Phi_0, \ \ \ \  V_{Xk}=(2n_k+1)e/C_k
\]
for all boxes
except $k=i$ and $j$. As shown in Fig.~3(b), the inductance $L$ is shared by the
Cooper-pair boxes $i$ and $j$ to form superconducting loops. The reduced Hamiltonian
of the system is given by~\cite{YTN}
\begin{equation}
H=\sum_{k=i,j}[\varepsilon_k(V_{Xk})\,\sigma_z^{(k)}-\overline{E}_{Jk}\;\sigma_x^{(k)}]
+\Pi_{ij}\,\sigma^{(i)}_x\sigma^{(j)}_x.
\end{equation}
Up to second-order terms,
\begin{equation}
\overline{E}_{Ji}(\Phi_{Xi},\Phi_e,L)\,=\,E_{Ji}(\Phi_{Xi})
\cos(\pi\Phi_e/\Phi_0)\,\xi,
\end{equation}
with
\begin{equation}
\xi=1-{1\over 2}(\eta_i^2+3\eta_j^2)\sin^2(\pi\Phi_e/\Phi_0),
\end{equation}
and
\begin{equation}
\Pi_{ij}=-LI_{ci}I_{cj}\sin^2(\pi\Phi_e/\Phi_0).
\end{equation}
Here the interbit coupling $\Pi_{ij}$ is controlled by both the
external flux $\Phi_e$ through the inductance $L$, and the local
fluxes, $\Phi_{Xi}$ and $\Phi_{Xj}$, through the SQUID loops.

Using these two types of circuits, we can derive the required one- and
two-bit operations for QC.
Specifically, the conditional gates such as the
controlled-phase-shift and controlled-NOT gates can be obtained
using one-bit rotations and only one basic two-bit operation. For
details, see Ref.~\cite{YTN}. A sequence of such conditional gates
supplemented with one-bit rotations constitute a universal element
for QC~\cite{LLOYD,DEU}. Usually, a two-bit operation is much
slower than a one-bit operation. Our designs for conditional gates
$U_{\rm CPS}$ and $U_{\rm CNOT}$ are {\it efficient} since {\it
only one} (instead of two or more) basic two-bit operation is
used.

\section*{Persistent currents and entanglement}

The one-bit circuit modeled by Hamiltonian (9) has two eigenvalues
$E^{(i)}_{\pm}=\pm E_i$,
with
\begin{equation}
E_i=[\varepsilon^2_i(V_{Xi})+
{\overline E}_{Ji}^2]^{1/2}.
\end{equation}
The corresponding eigenstates are
\begin{eqnarray}
|\psi^{(i)}_+\rangle&=&\cos\xi_i|\!\uparrow\rangle_i
-\sin\xi_i|\!\downarrow\rangle_i, \nonumber\\
|\psi^{(i)}_-\rangle&=&\sin\xi_i|\!\uparrow\rangle_i
+\cos\xi_i|\!\downarrow\rangle_i,
\end{eqnarray}
where
\begin{equation}
\xi_i={1\over
2}\tan^{-1}({\overline E}_{Ji}/\varepsilon_i).
\end{equation}
At these two
eigenstates, the persistent currents through the inductance $L$
are given by
\begin{equation}
\langle\psi^{(i)}_{\pm}|I|\psi^{(i)}_{\pm}\rangle=
\pm\left({{\overline E}_{Ji}I_{ci}\over E_i}\right)
\sin\left({\pi\Phi_e\over\Phi_0}\right)
+\left({\pi LI_{ci}^2\over 2\Phi_0}\right)
\sin\left({2\pi\Phi_e\over\Phi_0}\right),
\end{equation}
up to the linear term in $\eta_i$.
In the case when a dc SQUID magnetometer is inductively coupled to
the inductance $L$, these two supercurrents generate different
fluxes through the SQUID loop of the magnetometer and the
quantum-state information of the one-bit structure can be obtained
from the measurements.
In order to perform sensitive measurements with weak dephasing,
one could use the underdamped dc SQUID magnetometer designed
previously for the Josephson phase qubit~\cite{VAL}.

%\vspace{.2cm}\noindent

For the two-bit circuit described by Eq.~(15), the Hamiltonian has
four eigenstates and the supercurrents through inductance $L$ take
different values for these four states.
The fluxes produced by the supercurrents through $L$ can also be
detected by the dc SQUID magnetometer.
For example, when $\varepsilon_k(V_{Xk})=0$ and ${\overline
E}_{Jk}>0$ for $k=i$ and $j$, the four eigenstates of the two-bit
circuit are
\begin{eqnarray}
|1\rangle &=& {1\over 2}\left(|\uparrow\rangle_i|\uparrow\rangle_j
-|\uparrow\rangle_i|\downarrow\rangle_j
-|\downarrow\rangle_i|\uparrow\rangle_j
+|\downarrow\rangle_i|\downarrow\rangle_j\right),\nonumber\\
|2\rangle &=& {1\over 2}\left(|\uparrow\rangle_i|\uparrow\rangle_j
+|\uparrow\rangle_i|\downarrow\rangle_j
-|\downarrow\rangle_i|\uparrow\rangle_j
-|\downarrow\rangle_i|\downarrow\rangle_j\right),\nonumber\\
|3\rangle &=& {1\over 2}\left(|\uparrow\rangle_i|\uparrow\rangle_j
-|\uparrow\rangle_i|\downarrow\rangle_j
+|\downarrow\rangle_i|\uparrow\rangle_j
-|\downarrow\rangle_i|\downarrow\rangle_j\right),\nonumber\\
|4\rangle &=& {1\over 2}\left(|\uparrow\rangle_i|\uparrow\rangle_j
+|\uparrow\rangle_i|\downarrow\rangle_j
+|\downarrow\rangle_i|\uparrow\rangle_j
+|\downarrow\rangle_i|\downarrow\rangle_j\right).
\end{eqnarray}
Retained up to linear terms in
$\eta_i$ and $\eta_j$, the corresponding supercurrents through the
inductance $L$ are
\begin{equation}
\langle k|I|k\rangle={\cal I}_k\sin\left({\pi\Phi_e\over\Phi_0}\right)
+{\pi L{\cal I}_k^2\over 2\Phi_0}\sin\left({2\pi\Phi_e\over\Phi_0}\right)
\end{equation}
for $k=1$ to 4, where
\begin{eqnarray}
{\cal I}_1&=&-(I_{ci}+I_{cj}), \ \ \ \ \ \ \ \ {\cal I}_2=I_{cj}-I_{ci},
\nonumber\\
{\cal I}_3&=&I_{ci}-I_{cj}, \ \ \ \ \ \ \ \ {\cal I}_4=I_{ci}+I_{cj}.
\end{eqnarray}
These supercurrents produce different fluxes threading the SQUID
loop of the magnetometer and can be distinguished by dc SQUID
measurements.
When the two-bit system is prepared at the maximally entangled
Bell states
\begin{equation}
|\Psi^{(\pm)}\rangle={1\over\sqrt{2}}
(|\!\uparrow\rangle_i|\!\downarrow\rangle_j\pm|\!\downarrow\rangle_i
|\!\uparrow\rangle_j),
\end{equation}
the supercurrents through $L$ are given by
\begin{equation}
\langle\Psi^{(\pm)}|I|\Psi^{(\pm)}\rangle=
{\pi L\over 2\Phi_0}(I_{ci}\pm I_{cj})^2
\sin\left({2\pi\Phi_e\over\Phi_0}\right).
\end{equation}
These two states should be distinguishable by measuring the
fluxes, generated by the supercurrents, through the SQUID loop of
the magnetometer.

%\vspace{.2cm}\noindent

\section*{Estimates of the inductance for optimal coupling}

The typical switching time $\tau^{(1)}$ during a one-bit operation
is of the order of $\hbar/E_J^0$.
Using the experimental value $E_J^0\sim 100$~mK, then
$\tau^{(1)}\sim 0.1$~ns.
The switching time $\tau^{(2)}$ for the two-bit operation is
typically of the order
\[
\tau^{(2)} \sim (\hbar/L)(\Phi_0/\pi E_J^0)^2.
\]
Choosing $E_J^0\sim 100$~mK and
$\tau^{(2)}\sim 10\tau^{(1)}$ (i.e., ten times slower than the one-bit
rotation), we have
\[
L \sim 30~{\rm nH}
\]
in our proposal.  This number for $L$ is experimentally
realizable.
A small-size inductance with this value can be made with Josephson
junctions. Our expansion parameter $\eta$ is of the order
\[
\eta \sim \pi^2LE_J^0/\Phi_0^2\sim 0.1.
\]
Our inductance $L$ is related with the inductance $L'$ in ~\cite{MSS, ASZ} by
\begin{equation}
L'=(C_J/C_{qb})^2L.
\end{equation}
Let us now consider the case when $\tau^{(2)}\sim 10\tau^{(1)}$.
For the earlier design~\cite{ASZ}, $C_J\sim 11C_{qb}$ since $C_g/C_J\sim 0.1$,
which requires an inductance $L'\sim 3.6$~$\mu$H.
Such a large inductance is problematic to fabricate at nanometer
scales.
In the improved design~\cite{MSS}, $C_J\sim 2C_{qb}$, greatly
reducing the inductance to $L'\sim 120$ nH.  This inductance is
about four times larger than the one used in our scheme, making it
somewhat more difficult to realize than our proposed $L$.

\section*{Conclusion}

We propose a {\it scalable} quantum information processor with
Josephson charge qubits.
We use a common inductance to couple all charge qubits and design
{\it switchable\/} interbit couplings using two dc SQUIDs to
connect the island in each Cooper-pair box.
The proposed circuits are scalable in the sense that any two
charge qubits can be effectively coupled by an experimentally
accessible inductance.
In addition, we formulate \cite{YTN} an efficient QC scheme in
which only one two-bit operation is used in the conditional
transformations, including controlled-phase-shift and
controlled-NOT gates.

\section*{Acknowledgments}

We thank Yu.~Pashkin, B. Plourde and Xuedong Hu for useful discussions.
This work is supported in part by ARDA, the AFOSR, and the US National 
Science Foundation grant No.~EIA-0130383.

\end{article}
\end{document}